\documentclass[12pt,titlepage]{article}
\def\gtrsim{\mathrel{\raise.3ex\hbox{$>$}\mkern-14mu
             \lower0.6ex\hbox{$\sim$}}}
\def\ltorder{\mathrel{\raise.3ex\hbox{$<$}\mkern-14mu
             \lower0.6ex\hbox{$\sim$}}}

\def\dfrac#1#2{\displaystyle{\frac{{#1}}{{#2}}}}
\def\tag#1{}

\begin{document}
\begin{titlepage}
\begin{center}
{\LARGE Orbital Divergence and Relaxation in the Gravitational $N$-Body Problem}

\bigskip
{\Large P. Hut\footnotemark

and

D.C. Heggie\footnotemark

}

\end{center}

{\addtocounter{footnote}{-2}
\addtocounter{footnote}{1}\footnotetext{Institute for Advanced Study,
Princeton, 
U.S.A.}
\addtocounter{footnote}{1}\footnotetext{Dept. of Mathematics \& Statistics,
University of Edinburgh,
Edinburgh, 
U.K.
}
\vfill
{Suggested running head:  Orbital Divergence and Relaxation\\
\vfill
Address for Proofs:\\ 
Professor P. Hut,
Institute for Advanced Study,
Princeton,
NJ 08540,
U.S.A.\\
Phone: +1 609 734 8075, Fax: +1 609 924 8399, E-mail: piet@ias.edu}
\vfill
}
\end{titlepage}

\addtocounter{footnote}{-2}

\noindent \textbf{Abstract:} \ One of the fundamental aspects of
statistical behaviour in many-body systems is exponential divergence
of neighbouring orbits, which is often discussed in terms of Liapounov
exponents. \ Here we study this topic for the classical gravitational
$N$-body problem. \ The application we have in mind is to old stellar
systems such as globular star clusters, where $N\sim 10^{6}$, and so
we concentrate on spherical, centrally concentrated systems with total
energy $E<0.$ Hitherto no connection has been made between the time
scale for divergence (denoted here by $t_e$) and the time scale on
which the energies of the particles evolve because of two-body
encounters (i.e. the two-body relaxation time scale, $t_r$), even
though both may be calculated by similar considerations.

In this paper we give a simplified model showing that divergence in
phase space is initially roughly exponential, on a timescale
proportional to the crossing time (defined as a mean time for a star
to cross from one side of the system to another). In this phase
$t_e\ll t_r$, if $N$ is not too small (i.e. $N\gg30$). \ After several $e$-folding
times, the model shows that the divergence slows down. \ Thereafter
the divergence (measured by the energies of the bodies) varies with
time as $t^{1/2},$ on a timescale nearly proportional to the familiar
two-body relaxation timescale, i.e.  $t_e\sim t_r$ in this phase.
These conclusions are illustrated by numerical results.

\bigskip


\bigskip\bigskip

\noindent \textbf{1. \ Introduction}

\bigskip

The classical gravitational $N$-body problem is defined by the equations 
\begin{equation}
\mathbf{\ddot{r}}_{i}=-G\sum_{\begin{array}{c}j=1\\ j\neq i\end{array}}^{\displaystyle{N}}m_{j}\dfrac{%
\mathbf{r}_{i}-\mathbf{r}_{j}}{\left| \mathbf{r}_{i}-\mathbf{r}_{j}\right|
^{3}}  \tag{1}
\end{equation}
where $\mathbf{r}_{i}$ is the three-dimensional position vector of the $%
i^{th}$ star, $m_{i}$ is its mass, and $G$ is the universal constant of
gravitation.
We consider applications in which the total energy, $E,$ in the barycentric
frame is negative and the total angular momentum is negligible. \ Starting
from a rather broad set of initial conditions, such solutions settle down
into a roughly spherical distribution of bodies in approximate ``dynamic
equilibrium'' (Fig. 1), i.e. the spatial distribution is nearly
time-independent on the time scale of the orbital motions of the particles.

Early numerical integrations$^{[8]}$ with $N\leq 32$ showed that a
small change in initial conditions led to a roughly exponential divergence
of solutions (measured in $6N$-dimensional phase space), even though the
spatial distribution of the bodies in the two solutions might be
indistinguishable within statistical fluctuations. \ The timescale of
divergence, $t_{e}$, was of order the crossing time, $t_{cr},$ defined in a
certain conventional way as the time for a body with a typical speed to move
a distance comparable to the size, $R$, of the spatial distribution of the
particles$^{[1]}$.  Thus
$$
t_{cr}\sim {R\over V},\eqno(2)
$$
where $V$ is the root mean square speed of the particles. \ Later
work$^{[3,4,5]}$ extended numerical results to larger $N$, and Goodman
et al$^{[3]}$ devised theoretical models confirming that
$t_{e}/t_{cr}$ is virtually independent of $N.$ \ 

One particular
statistical specification of the initial conditions which has been
studied is the Plummer model, which is often used in stellar dynamics
for the study of relaxation and related processes.  It is the stellar
dynamical analogue of the $n=5$ polytrope.  For this model it has been
found$^{[4]}$ that
$$
t_{e}\simeq {0.116t_{cr}\over\ln \left(0.73 \ln N\right)}.
$$
The functional form is suggested by a theoretical model$^{[3]}$, 
and the coefficients are not thought to depend sensitively on the
initial conditions.  Therefore for large star clusters generally we have 
$$
t_e \sim 0.05t_{cr}.\eqno(3)
$$

\vspace*{3in}
\includegraphics{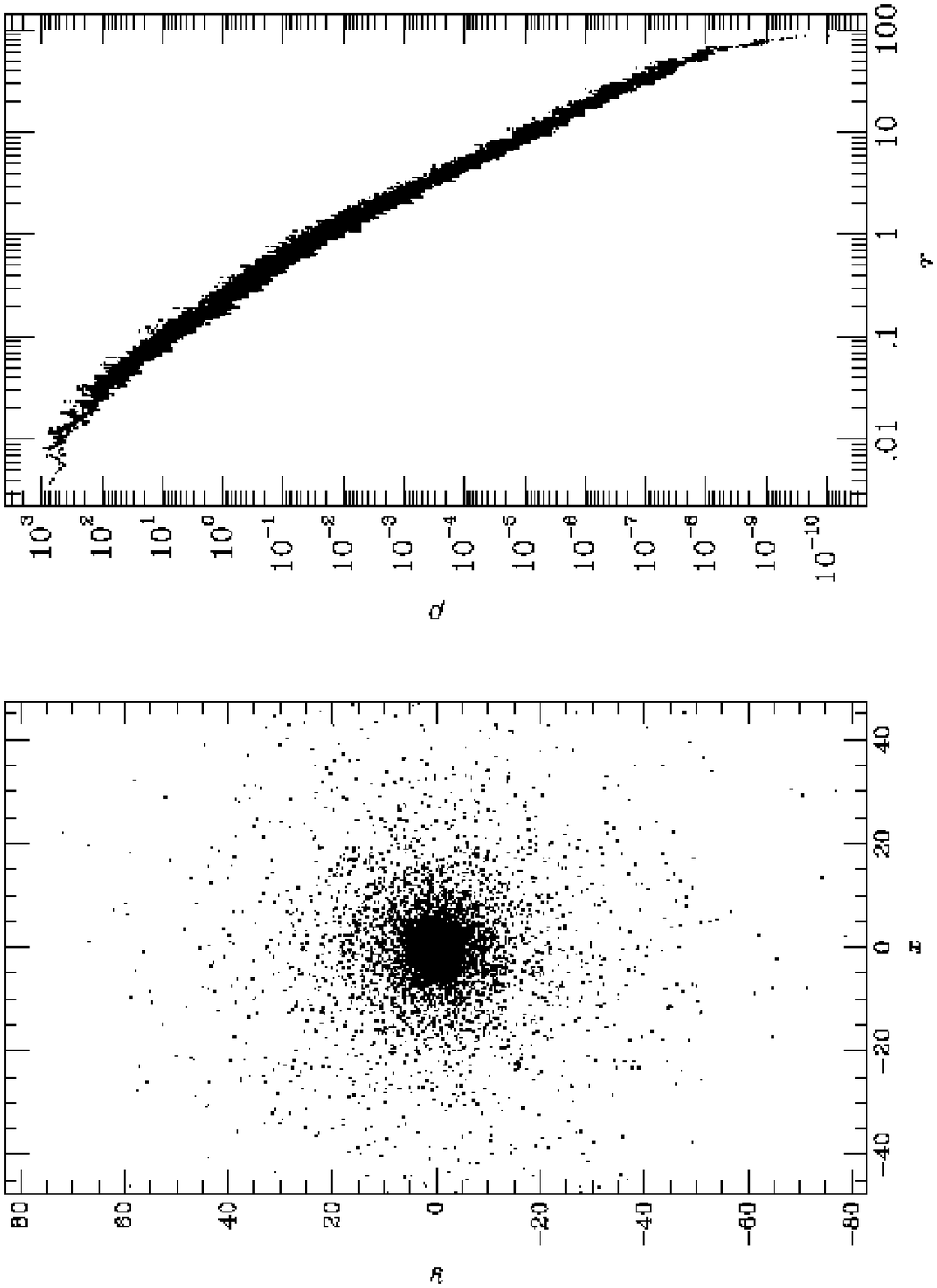}
\noindent Fig.1. Spatial distribution of bodies in a typical simulation.  On left
is a snapshot, and on the right is the numerically generated space
density as a function of radius.

 \bigskip

The theoretical models of Goodman et al$^{[3]}$ dealt with the linear divergence of
neighbouring solutions, when the separation in position of the $i^{th}$ body satisfies the
variational equation 
$$
\ddot{\delta \mathbf{r}_{i}}=-\sum_{\begin{array}{c}j=1\\ j\neq i\end{array}}^{\displaystyle{N}}m_{j}\left( 
\dfrac{\delta \mathbf{r}_{i}-\delta \mathbf{r}_{j}}{\left| \mathbf{r}_{i}-%
\mathbf{r}_{j}\right| ^{3}}-\dfrac{\left( \delta \mathbf{r}_{i}-\delta 
\mathbf{r}_{j}\right).\left( \mathbf{r}_{i}-
\mathbf{r}_{j}\right) }{\left| \mathbf{r}_{i}-\mathbf{r}_{j}\right|^5 }\left( 
\mathbf{r}_{i}-\mathbf{r}_{j}\right) \right) .\eqno(4)
$$
For practical purposes, however (e.g. for understanding the growth of errors
in a numerical integration) the resulting roughly exponential growth quickly leads to
separations so large that the linear approximation fails. \ In this contribution
we develop the simplest model of divergence to account for the later,
nonlinear growth of the separation between neighbouring solutions. \ We
shall see that the time dependence changes from roughly exponential to
roughly power-law, and that the timescale changes from roughly the  crossing time to nearly the two-body
relaxation timescale, $t_{r}.$ \ This is the timescale on which the energies
of the individual bodies vary significantly. \ Standard theory$^{[1,2]}$ shows that 
$$
t_{r}\sim \dfrac{N}{\ln N}t_{cr}\eqno(5)
$$
for systems
of the general kind considered here.

\bigskip

\noindent \textbf{2. \ A model of divergence}

\noindent 2.1 \ Linear growth of errors

In this section we introduce a toy model for the divergence of
neighbouring orbits.  Though it gives much insight into the physics of
the problem, many details are omitted.  In the first instance we apply
it to the linear regime in which the approximate eqs.(4) are valid.
In this regime more elaborate models have
been constructed by Goodman et al$^{[3]}$.

We make the following assumptions.  As in the theory of two-body
relaxation$^{[1,2]}$ we assume that the trajectory of a particle is
nearly rectilinear, except for occasional two-body encounters
(Fig. 2).  We suppose that the important encounters are in the
small-angle scattering regime, such that $p\gg Gm/v^2$ where $p$ is
the impact parameter and $v$ is the relative velocity of the two
particles.  In computing the effect of one encounter, we suppose we
can treat the scatterer as fixed.  We also suppose that successive
encounters can be treated as if motion takes place on one plane, and
that the difference between two orbits is measured by the difference
in the impact parameter, $\delta p$.  We assume that all particles
have the same mass $m$.  Finally, we suppose that the system is in virial
equilibrium (see Binney \& Tremaine$^1$), which implies that
$$
V^2 \sim
\dfrac{GmN}{R}.\eqno(6)
$$  
Here the symbol $\sim$ means ``is of order'',
i.e. that the relation is
approximate, and any numerical coefficient is ignored.  Thus $v\sim V$,
for example.

 \vspace*{2in}
 \includegraphics{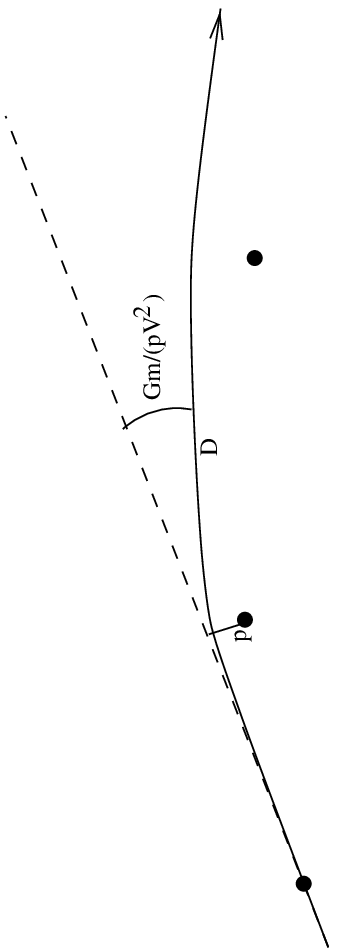}
\noindent Fig.2. Two successive encounters.

 \bigskip

In the small-angle scattering regime the maximum acceleration of the
moving particle is of order $\dfrac{Gm}{p^2}$ and the duration of the
encounter is of order $\dfrac{p}{V}$.  Thus the change in velocity is
of order $\dfrac{Gm}{pV}$, and so the angular deflection is of order
$\dfrac{Gm}{pV^{2}}$ (Fig.2). \ After the scattered body has travelled a
further distance $D$ to its next encounter, its spatial deflection is
of order
$\dfrac{GmD}{pV^{2}}$.

Now suppose the body had approached the first encounter on a parallel
path at a slightly different impact parameter $p+\delta p.$ \ Then, at
the time of the second encounter, its position would have been
displaced by a distance of order $\delta p+
\dfrac{GmD}{p^{2}V^{2}}\delta p.$ \ The first term is the displacement
that would have occurred even in the absence of the first encounter. \
The second occurs because, if $\delta p>0,$ the body has been
deflected less by the first encounter. \ (The differential
approximation used for this term is not valid unless $\vert\delta
p\vert\ll p;$ this is the approximation which restricts the present
theory to the linear regime in which eqs.(4) are valid.)\ The total displacement measures the
change in impact parameter at the second encounter. \ Hence the
variation in $p$ is multiplied by a factor of order $\left(
1+\dfrac{GmD}{p^{2}V^{2}}\right) $ per encounter\footnote{In a fully
three-dimensional treatment this becomes a matrix equation.}.

Now we consider the cumulative effect of several encounters within a
restricted range of impact parameters around the value $p$, e.g. from
$p/2$ to $2p$, but
ignoring other encounters.  We start at some time $t$ and consider the
effect of encounters in a subsequent interval $\Delta t$, chosen sufficiently large that several
encounters occur within this interval.  The actual number of such encounters is of order
$\dfrac{\Delta tV}{D},$ and so the
variation in the orbit is given\footnote{We ignore two complications which
tend to counteract each other: (i) the persistence of effects of early
encounters, and (ii) partial cancellation of successive encounters 
by their vectorial
character.} by 
\[
\delta r\left(t+\Delta t\right) \sim\delta r\left( t\right) \left( 1+\dfrac{GmD}{%
p^2V^2}\right) ^{\dfrac{\Delta tV}{D}}.
\]
Also it is clear that
$p^2Dn\sim1$, where $n$ is the number of particles per unit volume, and so
$$
\delta r\left(t+\Delta t\right) \sim\delta r\left( t\right) \left( 1+\dfrac{Gm}{%
p^{4}nV^{2}}\right) ^{\displaystyle{\Delta tVnp^{2}}}.  \eqno(7)
$$
It follows from the relation $n\sim\dfrac{N}{R^3}$ and eq.(6) that 
$$
\delta r\left(t+\Delta t\right) \sim\delta r\left( t\right) \left( 1+\dfrac{R^{4}}{%
p^{4}N^{2}}\right) ^{\dfrac{\Delta t}{t_{cr}}\dfrac{Np^{2}}{R^{2}}},  \eqno(8)
$$
where we have used eq.(2).

Encounters take place at a wide range of impact parameters $p.$ \ Writing
eq. (8) as 
$$
\ln \delta r\left(t+\Delta t\right) - \ln \delta r\left( t\right) \sim
\dfrac{\Delta t}{%
t_{cr}}N\dfrac{p^{2}}{R^{2}}\ln \left( 1+\dfrac{R^{4}}{p^{4}N^{2}}\right)  
\eqno(9)
$$
we see that those with $p<<RN^{-1/2}$ are individually very effective but
too rare to dominate, whereas those with $p>>RN^{-1/2}$ lose out by being
individually ineffective, despite being very numerous. \ Encounters at
impact parameter $p\sim RN^{-1/2}$ are most effective cumulatively, and lead
to exponential growth of the deviation $\delta r,$ on a timescale of order $%
t_{cr}$.  

Another way of seeing this is to sum the right hand side of eq.(9) over
all impact parameters $p\ltorder R$.  Since this term represents the
effect of encounters with impact parameters in a range near some value
$p$, the summation can be accomplished by multiplying by $dp/p$ and
integrating.  The result is that
$$
\ln \delta r\left(t+\Delta t\right) - \ln \delta r\left( t\right) \sim
\dfrac{\Delta t}{%
t_{cr}}\ln N.
$$
Except for the logarithmic factor, this is equivalent to the result
obtained by ignoring all encounters except those near $p\sim RN^{-1/2}$.

Many factors have been omitted from this simple model, including the
distribution of velocities and density, and the curved orbits of bodies
between encounters. \ Nevertheless, the results of more detailed models and
numerical simulations, already quoted, confirm our basic result, except for
a very weak $N$-dependence.

\bigskip

\noindent 2.2 \ Nonlinear growth of separation

The above theory is valid as long as $\delta r<<p,$ and here we may take for 
$p$ the impact parameter for the most effective encounters, i.e. $p\sim
RN^{-1/2}.$ \ Suppose we are interested in growth of errors in an $N$-body
integration of eqs. (1), for a system which has been scaled so that $R\sim
1.$ \ Then we may have $\delta r(0)\sim 10^{-16}$ for a double precision
calculation, and so the linear approximation breaks down after about
$30t_e$, i.e. between one and two  $t_{cr}$, by eq.(3).

Thereafter we suppose that encounters with impact parameters $p<<\delta r$
are ineffective. \ Then we may estimate the growth of the separation of
neighbouring orbits by substituting $p\sim \delta r(t)$ in eq. (9), which gives
\[
\ln \delta r(t+\Delta t)- \ln \delta r\left( t\right)\sim
\dfrac{\Delta t}{t_{cr}}\dfrac{%
N \delta r(t) ^{2}}{R^{2}}\ln \left( 1+\dfrac{R^{4}}{N^{2}
\delta r(t) ^{4}}\right).
\]
We are
in a regime where $\delta r(t)\gtrsim RN^{-1/2},$ and so we can approximate
$$
\ln \delta r(t+\Delta t)-\ln \delta r(t) \sim\dfrac{\Delta t}{t_{cr}}\dfrac{R^{2}%
}{N\delta r(t)^{2}}\;.  \eqno(10)
$$
Since  the term on the right depends on $t$, we
can no longer conclude that 
$\ln \delta r(t)$  increases linearly
with $t.$ \ To determine its time dependence we rewrite eq. (10) as a
differential equation, i.e.
\[
\dfrac{d}{dt}\ln \delta r(t)\sim\dfrac{1}{t_{cr}}\dfrac{R^{2}}{N\delta
r(t)^{2}}\;.
\]
Ignoring for the moment the distinction between ``$\sim$'' and
``$=$'', we obtain the solution 
\[
\delta r\left( t\right) =\left(  \delta r\left(t_0\right) 
^{2}+2\dfrac{t-t_0}{t_{cr}}\dfrac{R^{2}}{N}\right) ^{1/2},
\]
where $t_0$ is a constant, which may be interpreted as the time at
which the growth of errors enters the nonlinear regime.

Well into the nonlinear regime we now see that $\delta r\left( t\right)
\sim R\left( \dfrac{t}{Nt_{cr}}\right) ^{1/2}.$ \ In order to interpret this result we shall estimate the difference
in binding energy, $\varepsilon,$ of the body between the two
neighbouring solutions. \ Now $\varepsilon\sim \dfrac{GNm}{R},$ and we
can estimate $\delta\varepsilon\sim \dfrac{GNm\delta r}{R^{2}}.$   (We
could obtain a similar estimate from consideration of the difference
in velocity.)\ Hence $\dfrac{%
\delta\varepsilon}{\varepsilon}\sim \left( \dfrac{t}{Nt_{cr}}\right) ^{1/2}.$ \ Now the
two-body relaxation time, $t_{r}$, may be estimated by eq.(5), and so $\dfrac{\delta\varepsilon}{\varepsilon}\sim \left( \dfrac{t%
}{t_{r}}\right) ^{1/2}$ if we ignore a logarithmic dependence on $N.$

\bigskip 

\noindent \textbf{3. \ Discussion}

Recall that we are considering two solutions of eq. (1) starting with
slightly different initial conditions.  Suppose that we measure the
separation of the two solutions by the separation in energy, $\delta
\varepsilon,$ of a typical body. \ What we have concluded is that, for
at most a few crossing
times, $\delta \varepsilon(t)$ grows exponentially, with an $e$-folding time comparable
with $t_{cr}$ itself. \ Thereafter $\delta \varepsilon(t)$ approaches a power law
dependence, varying as $t^{1/2},$ on a timescale of the relaxation time.

The standard theory of relaxation tells us how $\varepsilon$ (the energy of a given
star) evolves on a {\sl single} solution of the $N$-body equation. \ If we
ignore variations of $\varepsilon$ inside an encounter, $\varepsilon$ performs a random walk on
the timescale $t_{r},$ and the change in $\varepsilon$ varies as $t^{1/2}.$ \ (We here
ignore the role of ``dynamical friction'', which corresponds to the drift
term in a Fokker-Planck description of the relaxation$^{[1,2]}$.)

Fig.3 illustrates these points using data from numerical $N$-body
integrations with $N = 256$.  Two systems were integrated
simultaneously using identical initial conditions except for a small
difference in one coordinate of one particle.  The solid curve (a)
shows the mean square difference in the energies of the $N$
particles\footnote{Similar results have been presented by
Merritt$^{[7]}$ for motion in the gravitational field of $N$ {\sl
fixed} bodies}.
The corresponding initial conditions were also used for simultaneous
integration of the variational equations, and the long-dashed curve
(b) shows the corresponding mean square variation of energy.  This
grows nearly exponentially, but is followed by (a) for only a limited
time of order a crossing time.  The short-dashed curve (c) shows the
mean square difference between the initial energy and the energy at
time $t$, again averaged over the $N$ particles.  This is caused by
two-body relaxation.  Evidently curve (a) departs from curve (b)
around the point where the latter crosses curve (c), and then nearly
follows (c).
In this way we see that the
growth of errors, which is exponential only in the linear regime, is
consistent with the theory of two-body relaxation.

\vspace*{4in}
\includegraphics{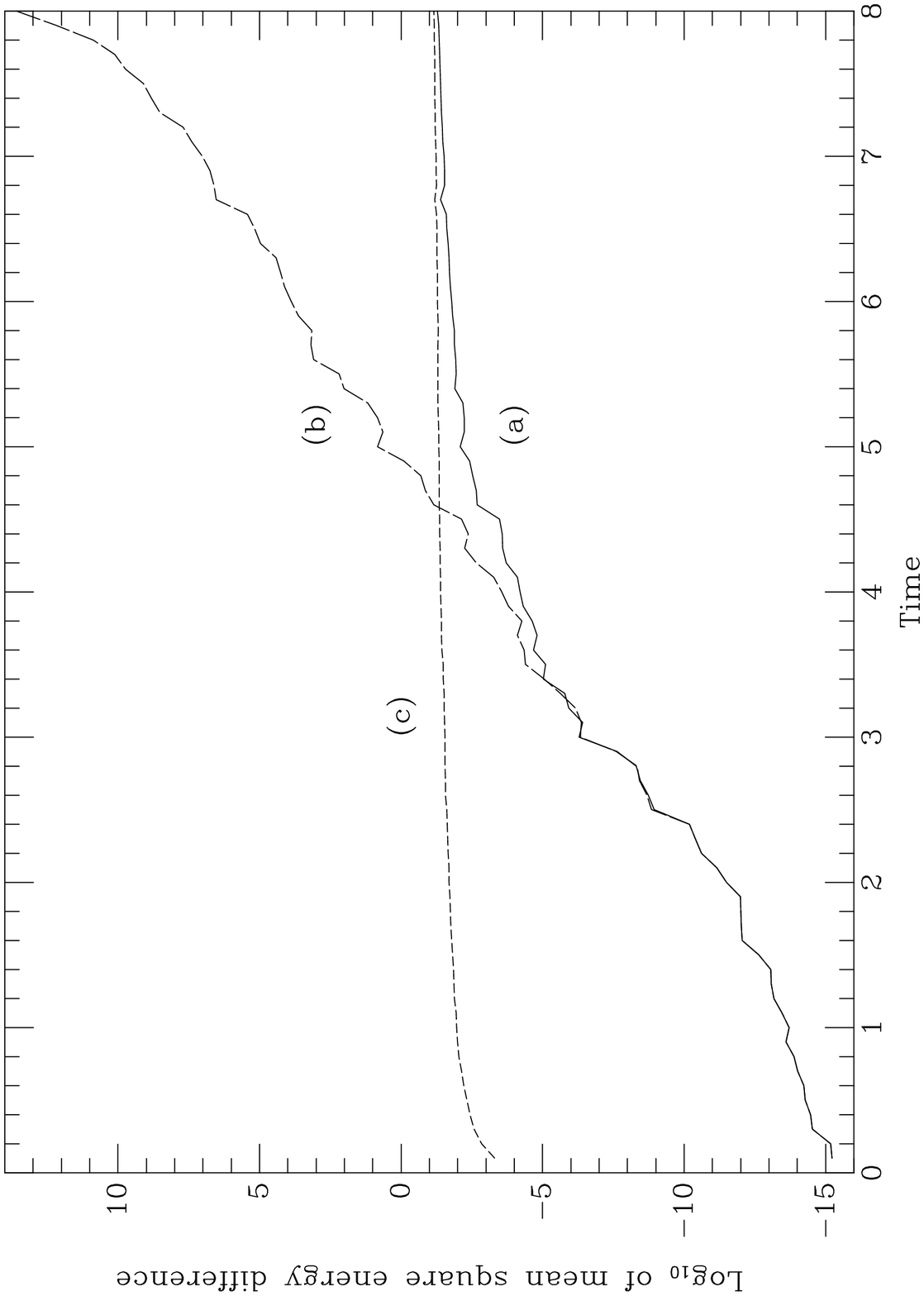}
Fig.3. Mean square energy difference in numerical integrations with
$N=256$, as a function of time.  The meaning of the different curves
is stated in the text.  The results plotted are the mean of four
independent runs.  In the adopted units the crossing time is
$2\sqrt{2}$.
\bigskip

The exponential divergence slows down to a power-law growth because
close encounters become increasingly ineffective.  There is a
geometric way of looking at this.  Krylov$^{[6]}$ showed that the
divergence could be understood as the behaviour of neighbouring
geodesics on a certain manifold.  As two neighbouring geodesics
deviate further, their deviation is influenced less and less by the
fine geometrical structure of the manifold across which they are
proceeding.

\bigskip 
\noindent{\bf Acknowledgement} We thank the referees for their comments
and suggestions.

\bigskip\noindent{\bf References}

{\leftskip=\parindent
\parindent=-\leftskip

1. Binney J., Tremaine S. (1987).  {\sl Galactic Dynamics} (Princeton
University Press, Princeton)

2. Chandrasekhar S., (1942). {\sl Principles of Stellar Dynamics}
   (University of Chicago Press, Chicago; also Dover Publications, New
   York [1960])

3. Goodman J., Heggie D.C., Hut P. (1993). On the Exponential Instability
of $N$-Body Systems, {\sl ApJ}, {\bf 415}, 715-33 

4.  Hemsendorf M., Merritt D. (2002).  Instability of the
Gravitational $N$-Body Problem in the Large-$N$ Limit, astro-ph/0205538

5. Kandrup H.E., Smith H. (1991).  On the sensitivity of the $N$-body
problem to small changes in initial conditions, {\sl ApJ}, {\bf 374},
255-65

6. Krylov N.S. (1979).  {\sl Works on the Foundations of Statistical
Physics} (Princeton University Press, Princeton)

7. Merritt D. (2001). Non-integrable Galactic Dynamics, in
   B.A. Steves, A.J. Maciejewski, eds, {\sl The Restless Universe},
   (SUSSP Publications, Edinburgh; and Institute of Physics
   Publishing, London)

8. Miller R.H. (1964).  Irreversibility in Small Stellar Dynamical
Systems, {\sl ApJ}, {\bf 140}, 250-6

}



\end{document}